\documentclass[floatfix,
  aps,
  pra,
  reprint,
  superscriptaddress,
  nofootinbib
]{revtex4-2}

\usepackage[T1]{fontenc}
\usepackage{lmodern}

\usepackage{amsmath}
\usepackage{amssymb}
\usepackage{amsfonts}
\usepackage{amsthm}
\usepackage{bm}
\usepackage{mathtools}

\usepackage{graphicx}
\usepackage{xcolor}
\usepackage{booktabs}
\usepackage{microtype}

\usepackage[
  colorlinks=true,
  linkcolor=blue,
  citecolor=blue,
  urlcolor=blue
]{hyperref}

\numberwithin{equation}{section}

\newtheorem{theorem}{Theorem}[section]

\theoremstyle{definition}
\newtheorem{definition}[theorem]{Definition}

\theoremstyle{remark}

\newcommand{\code}[1]{\texttt{\nolinkurl{#1}}}

\DeclareMathOperator*{\argmax}{arg\,max} 

\begin{document}

\title{Efficient \texorpdfstring{$XX$}{XX} Placement Strategies for Energy-Gap Enhancement in Quantum Annealing}

\author{Jennivine Chen}
\email{j.chen@liacs.leidenuniv.nl}
\affiliation{
Department of Physics and Astronomy,
University College London,
WC1E 1BT London, UK
}

\author{Natasha Feinstein}
\email{natasha.feinstein.19@ucl.ac.uk}
\affiliation{
London Centre for Nanotechnology,
University College London,
WC1H 0AH London, UK
}

\author{P. A. Warburton}
\email{p.warburton@ucl.ac.uk}
\affiliation{
London Centre for Nanotechnology,
University College London,
WC1H 0AH London, UK
}
\affiliation{
Department of Electronic \& Electrical Engineering, 
University College London,
WC1E 7JE London, UK
}

\begin{abstract}
Quantum annealing offers the prospect of a computational advantage over classical methods for certain classes of combinatorial optimisation problems. Its performance, however, is fundamentally constrained by the minimum spectral gap between the ground state and the first excited state encountered during the annealing process. When this gap closes exponentially with problem size, the runtime required to maintain adiabatic evolution can likewise grow exponentially, severely limiting the practical efficiency of the annealer. For problems including the maximum weight independent set (MWIS), previous work has shown that introducing $XX$ interactions along problem-graph edges can substantially enlarge the minimum energy-gap and thereby improve annealing performance. In practice, however, current quantum annealing hardware does not support these interactions natively and methods for implementing effective $XX$ couplings come with hardware overhead and connectivity constraints. These limitations motivate the use of only a limited number of $XX$ interactions. In this paper, we investigate $XX$ placement strategies for MWIS instances under both resource and topological constraints, and demonstrate that careful selection of the placements can substantially increase the minimum energy-gap. We evaluate these strategies through a detailed statistical analysis of both deliberately constructed hard MWIS instances and random Erd\H{o}s--R\'enyi graphs. Notably, these strategies are particularly effective for hard MWIS instances with initially small minimum gaps, where targeted placement yields the largest relative improvements.
\end{abstract}

\maketitle


\section{Introduction}
Combinatorial optimisation problems manifest practically in areas such as routing \cite{sales2023}, resource allocation \cite{gabbassov2022, witt2023, otsuka2022}, and portfolio optimisation \cite{rubiogarcia2022}. Rather than applying classical optimisation algorithms, quantum annealing (QA) provides a framework for solving combinatorial optimisation problems by encoding the solution to a problem instance in the ground state of a problem Hamiltonian. 

By applying QA, the hope is for quantum effects, in particular tunnelling, to allow the system to move through optimisation landscapes in ways that differ from classical thermal search, therefore making the search more efficient~\cite{santoro2002}. Starting from the ground state of an initial Hamiltonian that is easy to prepare, the system is evolved continuously towards the problem Hamiltonian. Provided that the evolution is sufficiently slow, the system remains in its instantaneous ground state and therefore terminates in a state encoding the optimal solution \cite{farhi2000,albash2018}. 

Unfortunately QA is limited by the spectral gap between the instantaneous ground state and the excited states. The runtime required for adiabatic evolution scales inversely with the square of the minimum instantaneous spectral gap during the anneal; consequently, instances for which the gap becomes exponentially small with increasing problem size can require exponentially long annealing times. Avoided level crossings associated with small gaps therefore constitute an important bottleneck to the practical performance of QA.

One approach to mitigating this problem is to modify the annealing path by introducing additional terms into the annealing Hamiltonian. There has been a particular interest in the introduction of two-local $XX$ interactions, which have been shown to substantially enlarge the minimum gap for various QA settings \cite{SekiYuyaandNishimori2012,Hormozi2017,Albash2019,Mehta2021,feinstein2024-xx-catalyst,ghosh2024}. In the context of the maximum weight independent set (MWIS) problem, previous work has demonstrated that placing $XX$ interactions along edges of the problem graph can significantly improve the minimum spectral gap \cite{nutricati2025}.

The unrestricted addition of $XX$ interactions, however, is unlikely to be practical on near-term QA hardware. Currently, neither superconducting nor neutral-atom analogue platforms on which QA is run natively support two-local $XX$ interactions. For superconducting flux qubits of the type used by D-Wave, for example, an $XX$ interaction could theoretically be built directly into hardware via inductive \cite{zhang2023tunableinductivecouplerhigh} or capacitive \cite{PhysRevApplied.13.034037} couplers. However, as Kerman has shown \cite{Kerman_2019}, both methods only yield a significant $XX$ interaction when large local $X$ fields are applied to both qubits. Gadgets that implement \textit{effective} $XX$ terms \cite{banks2025,banks2026,Headley_2026} using available interactions come with their own hardware overhead and connectivity restrictions. This motivates a more constrained question: \emph{if only a limited number of $XX$ edges can be placed, where should they be positioned in order to obtain the greatest improvement in the minimum gap?}

In this work, we investigate this question for MWIS instances under both resource and connectivity constraints. Rather than assuming arbitrary native $XX$ interactions, we consider specifically the constraints associated with the parity-encoding based gadget proposed in \cite{banks2026}. The gadget is able to implement \textit{chains} of effective $XX$ interactions with an overhead of one additional qubit per chain. Experimental implementation of the gadget on D-Wave hardware yielded a three order of magnitude ground-state fidelity improvement for an illustrative setting. However, in addition to being topologically restricted to chains of interactions, the gadget comes with other subtleties that encourage limiting the number of $XX$ interactions to be introduced. We therefore study $XX$ placement as a constrained selection problem, in which both the number and topology of the available $XX$ interactions are restricted.

We compare several strategies for selecting coupling locations and evaluate their performance on both deliberately constructed hard MWIS instances and ensembles of Erd\H{o}s--R\'enyi (ER) random graphs. Our numerical results show that strategically chosen subsets of $XX$ edges can produce substantial enhancements of the minimum energy gap despite these implementation constraints. Notably, the improvement is most pronounced for instances where the original annealing paths exhibit particularly small minimum gaps, suggesting that selective coupling placement is especially useful for the hardest instances.

The remainder of this paper is organised as follows. Section~\ref{sec:quantum_annealing_MWIS} introduces the MWIS problem and its encoding as a quantum annealing Hamiltonian. Section~\ref{sec:ancilla_implementation} describes the parity-based gadget implementation of effective $XX$ couplings and the topological constraints that it imposes. Section~\ref{sec:strategies} introduces the $XX$-interaction placement strategies. Sections~\ref{sec:pseudo-hard_MWIS} and~\ref{sec:ER_instances} present our numerical analysis on planted hard- and random ER-instances of the MWIS problem. We conclude in Sec.~\ref{sec:conclusion} with a discussion of the implications of these results and possible extensions.

\section{Quantum annealing for the MWIS problem}
\label{sec:quantum_annealing_MWIS}
Let
\begin{equation}
    G=(V,E)
\end{equation}
be an undirected graph with vertex set $V=\{1,\ldots,n\}$ and edge set $E$, and let each vertex $i\in V$ be assigned a positive weight $w_i$. An \emph{independent set} is a subset $S\subseteq V$ such that no two vertices in $S$ are adjacent. The maximum weight independent set problem asks for an independent set of maximum total weight,
\begin{equation}
    S^\star
    =
    \argmax_{\substack{S\subseteq V\\S\ {\rm independent}}}
    \sum_{i\in S} w_i.
\end{equation}

By introducing binary variables $x_i\in\{0,1\}$, where $x_i=1$ indicates that vertex $i$ is included in the independent set, MWIS can be expressed as the constrained optimisation problem
\begin{equation}
    \max_{\bm{x}\in\{0,1\}^n}
    \sum_{i=1}^{n} w_i x_i,
\end{equation}
subject to
\begin{equation}
    x_i x_j = 0,
    \qquad
    (i,j)\in E.
\end{equation}

To encode this as a quantum Hamiltonian, we make the standard translation
\begin{equation}
    x_i = \frac{1+z_i}{2},
\end{equation}
where $z_i = \pm 1$. So, provided the edge penalty is chosen sufficiently large, the ground state of the Hamiltonian 
\begin{equation}
    \hat{H}_p = \sum_{i \in V} (d_iJ_{zz} - 2w_i)Z_i + \sum_{(i,j) \in E} J_{zz} Z_i Z_j
    \label{eq:problem_hamiltonian}
\end{equation}
corresponds to a maximum-weight independent set of the original graph. In Eq.~\eqref{eq:problem_hamiltonian}, $d_i$ is the degree of vertex $i$, and $w_i$ is its weight.

The conventional transverse-field driver Hamiltonian is
\begin{equation}
    H_{\mathrm{D}}
    =
    -\sum_{i=1}^{n} X_i,
    \label{eq:driver_hamiltonian}
\end{equation}
whose ground state is the uniform superposition of computational basis states. A standard annealing schedule interpolates between Eqs.~\eqref{eq:driver_hamiltonian} and \eqref{eq:problem_hamiltonian},
\begin{equation}
    H(s)
    =
    A(s)H_{\mathrm{D}}
    +
    B(s)H_{\mathrm{P}},
    \qquad
    s\in[0,1],
    \label{eq:standard_anneal}
\end{equation}
where $A(0)\gg B(0)$ and $B(1)\gg A(1)$. Furthermore, the spectral gap of the Hamiltonian is
\begin{equation}
    \Delta(s) = E_1(s)-E_0(s),
\end{equation}
and its minimum over the anneal is defined as
\begin{equation}
    \Delta_{\min}
    =
    \min_{s\in[0,1]}\Delta(s).
\end{equation}
For difficult instances, the interpolation in Eq.~\eqref{eq:standard_anneal} may encounter a narrow avoided crossing and hence a small value of $\Delta_{\min}$. It is these small gaps which we aim to mitigate through the introduction of $XX$ interactions.

\section{Parity-based gadget implementation of \texorpdfstring{$XX$}{XX} couplings}
\label{sec:ancilla_implementation}

To obtain a hardware-motivated model for how such $XX$ interactions might enter the anneal in practice, we consider the parity-based $XX$ gadget introduced in~\cite{banks2026}. We summarise its key concepts below and refer interested readers to that work for the full description.

The purpose of the gadget is to provide an \textit{encoded logical} setting that includes $XX$ interactions with a \textit{physical} setting that only includes fields and interactions that we have access to -- \textit{i.e.}, $X$ fields, $Z$ fields and $ZZ$ interactions. To implement an $XX$ interaction between two logical qubits, we encode each logical qubit in the parity of a pair of physical qubits, with the two pairs sharing one physical qubit. Denoting logical variables by a tilde, the encoding is defined as
\begin{equation}
    \tilde z_1 = z_0 z_1,
    \qquad
    \tilde z_2 = z_1 z_2.
\end{equation}
The corresponding logical Pauli-$Z$ operators are therefore
\begin{equation}
    \widetilde Z_1 = Z_0 Z_1,
    \qquad
    \widetilde Z_2 = Z_1 Z_2.
\end{equation}
Since $Z^2=I$, a $ZZ$ interaction between the two logical qubits reduces to a two-body interaction between the outer physical qubits:
\begin{equation}
    \widetilde Z_1\widetilde Z_2
    = Z_0 Z_1 Z_1 Z_2
    = Z_0 Z_2.
\end{equation}

The logical transverse operators can be identified by considering how flipping each physical qubit affects the encoded parities. Flipping either outer qubit, $q_0$ or $q_2$, changes only one of the two logical parities. Thus, the individual logical $X$ operators can be represented as
\begin{equation}
    \widetilde X_1 = X_0,
    \qquad
    \widetilde X_2 = X_2.
\end{equation}
In contrast, the central physical qubit $q_1$ is shared by both parity pairs, so flipping it simultaneously reverses both logical parities. Its transverse operator therefore implements the logical $XX$ interaction:
\begin{equation}
    \widetilde X_1\widetilde X_2 = X_1.
\end{equation}

The gadget can be extended to produce a chain of logical qubits with $XX$ interactions between them through effectively chaining parity encoded qubits together. The cost is the loss of local $X$ driving on the internal logical qubits. This can be understood through the fact that neither of the two physical qubits encoding these internal logical qubits is associated with that logical qubit alone.

The parity-based gadget construction also imposes restrictions on the placement of the $XX$ terms. Under the parity encoding, a transverse field acting on a physical qubit induces a transverse $\widetilde X_i$ operator on each parity variable involving that qubit. Consequently, a physical qubit belonging to a single parity variable gives rise to a one-body transverse term, whereas a qubit shared by two parity variables generates a two-body $XX$ interaction. More generally, a branching point of degree $k$ would produce a $k$-body transverse interaction in the parity-encoded description. To retain a driver Hamiltonian containing only one- and two-body transverse terms, nodes in the $XX$-interaction graph must therefore have maximum degree two. Its connected components are consequently restricted to non-branching chains or closed loops.

The gadget comes with one additional complication when embedded within a larger graph structure. The parity encoding of a logical qubit into two physical qubits means that logical $ZZ$ term between an encoded qubit and a non-encoded qubit require physical three-local $Z$ interactions. Similarly,  a $ZZ$ interaction between two encoded qubits that are not adjacent in the $XX$-interaction graph requires a physical four-local $Z$ term. Such three- and four-local $Z$ interactions can be emulated through standard reduction techniques at the cost of one and two additional ancillary qubits respectively per interaction \cite{dattani2019quadratizationdiscreteoptimizationquantum}. This additional qubit overhead motivates limiting the number of gadget $XX$ interactions that are included.

In this work we will not be considering the physical Hamiltonian associated with the gadget. Rather, we directly introducing $XX$ interactions into the annealing Hamiltonian in a way that is \textit{guided} by the constraints on the $XX$-interaction graph that the gadget imposes. The $XX$ interactions enter the annealing Hamiltonian as part of the driver, such that the new driver can be written as
\begin{equation}
     H_{\mathrm{D}} = H'_{\mathrm{D}} -\sum_{(i,j) \in E_{\mathrm{xx}}}^{n}  X_i X_j,
     \label{eq:new_driver}
\end{equation}
where $E_{\mathrm{xx}}$ denotes the set of $XX$ edges and is restricted to consisting of disconnected chains and loops, and $H'_{\mathrm{D}}$ is obtained from standard driver by removing the single-qubit $X$ fields on qubits incident to two edges in $E_{\mathrm{xx}}$.

The practical placement problem studied in the following section can therefore be formulated schematically as
\begin{equation}
    \underset{E_{\mathrm{xx}}}{\text{choose}}
    \quad
    E_{\mathrm{xx}}
    \subseteq E
\end{equation}
so as to improve
\begin{equation}
    \Delta_{\min}
    \bigl(
        E_{\mathrm{xx}}
    \bigr),
\end{equation}
subject simultaneously to the $XX$-budget constraint
$|E_{\mathrm{xx}}|\leq K$ and the topological constraints
arising from the parity-based gadget implementation. Section~\ref{sec:strategies} considers several strategies for making this choice, while their numeral analysis on planted hard- and ER-instances can be found in Section~\ref{sec:pseudo-hard_MWIS} and~\ref{sec:ER_instances}.

\section{Constrained \texorpdfstring{$XX$}{XX} placement strategies}
\label{sec:strategies}
In our investigation, we consider the following $XX$ placement strategies, each inspired by some property of the MWIS problem instance. All three strategies can be created prior to running the quantum annealer, in time polynomial in the size of the graph.

\begin{definition}
    (Vertex weight conflict). For MWIS, difficult competition often occurs around edges connecting two high-weight vertices, where both vertices are individually attractive, but they cannot both be selected since that would violate the independent-set constraint. 
    
    This strategy therefore ranks each problem edge $(i,j)$ by its score 
    \begin{equation}
        \min(w_i, w_j),
    \end{equation}
    in descending order, and selects the top $k$ edges with the highest $\min(w_i, w_j)$.

\end{definition}

\begin{definition}
    (Field-strength ambivalence). The problem Hamiltonian expressing the MWIS problem as equation~\eqref{eq:problem_hamiltonian} encodes the tension when choosing whether to include a vertex $i$ in the independent-set through the term 
    \begin{equation}
        h_i = (d_iJ_{zz} - 2w_i)Z_i.
    \end{equation}
    On the one hand, since $z_i = \pm 1$, including vertex $i$ in the set will lower the energy of the Hamiltonian by $4w_i$, but at the same time it raises the energy by the penalty term $d_iJ_{zz}$. Therefore, we reason that vertices with small $h_i$ values are close to being locally balanced between selected and not selected, and are likely to be sites of ambivalence. 

    Thus, this strategy ranks each problem edge $(i,j)$ by its score 
    \begin{equation}
        s = \frac{1}{|h_i| + |h_j|},
    \end{equation}
    in descending order, and outputs the top $k$ edges with the highest score $s$.
\end{definition}

\begin{definition}
    (Edge-betweenness centrality). In graph theory, centrality quantifies the structural importance or influence of a vertex or edge within a graph. There are various measures of centrality, but for our purposes we will use the measure of shortest-path edge-betweenness implemented by python's \code{NetworkX} package~\cite{hagberg2008,networkx_centrality_docs}.

    The shortest-path edge-betweenness centrality measures how often an edge lies on shortest paths between other pairs of vertices. An edge has high betweenness if it acts as a bottleneck or bridge through which many graph paths pass. For an edge $e$, is it defined as
    \begin{equation}
        C_B(e)=\sum_{s,t \in V}\frac{\sigma(s, t | e)}{\sigma(s, t)},
    \end{equation}
    where $V$ is the set of vertices, $\sigma(s,t)$ is the number of shortest-paths between vertex $s$ and $t$, and $\sigma(s,t | e)$ is the number of those paths passing through edge $e$.
    
    For a graph $G = (V, E)$, the function \code{NetworkX.edge_betweenness_centrality(G)} computes this quantity efficiently in $O(|V|\cdot|E|)$ time using Brandes' algorithm~\cite{BRANDES2008136}. In this strategy, we therefore rank each problem edge $(i,j)$ by its edge-betweenness centrality score, in descending order. Then the strategy produces $k$ edges with the highest shortest-path betweenness score.
\end{definition}

\section{Numerical analysis on constructed small-gap instances}
\label{sec:pseudo-hard_MWIS}
We follow the procedure introduced in~\cite{feinstein2024} to generate $100$ pseudo-random $10$-vertex MWIS graphs with small minimum spectral gaps and four planted optima per graph. The vertices comprising each independent set (i.e., local optimum) are selected probabilistically, with vertices already shared by another set suppressed by a factor of $0.5$. The global optimum is assigned a total weight of $2.05$, while the remaining planted optima are assigned weights drawn uniformly from the interval $[1.9,2.00]$, thereby enforcing a minimum weight separation of $0.05$. Then the remaining weight associated with each planted optimum is distributed evenly among its constituent vertices. Edges are then introduced between pairs of vertices that do not belong to a common planted optimum.

\begin{figure}[t]
    \centering
    \includegraphics[width=0.85\linewidth]{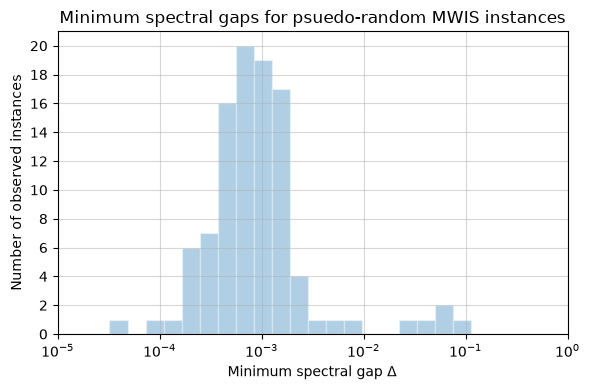}
    \caption{Distribution of minimum spectral gaps, $\Delta$, for MWIS instances generated using the pseudo-random construction prior to the introduction of $XX$ interactions. The histogram shows the number of instances within logarithmically spaced bins. The distribution is concentrated between $10^{-4} < \Delta \leq 10^{-2}$, indicating that the construction predominantly generates small-gap instances.}
    \label{fig:distribution_of_delta}
\end{figure}

For MWIS problem instances generated using the pseudo-random construction described above, Fig.~\ref{fig:distribution_of_delta} shows the distribution of minimum spectral gaps, $\Delta$, prior to the introduction of $XX$ interactions. The predominance of small gaps is particularly relevant to our investigation, as these instances present the greatest potential for improvement through $XX$ terms .

Now we present the numerical results analysing the effect of the different $XX$ placement strategies on these pseudo-random MWIS instances.

\subsection{Comparison of \texorpdfstring{$XX$}{XX} placement strategies}
\begin{figure*}[t]
    \centering
    \includegraphics[width=0.95\linewidth]{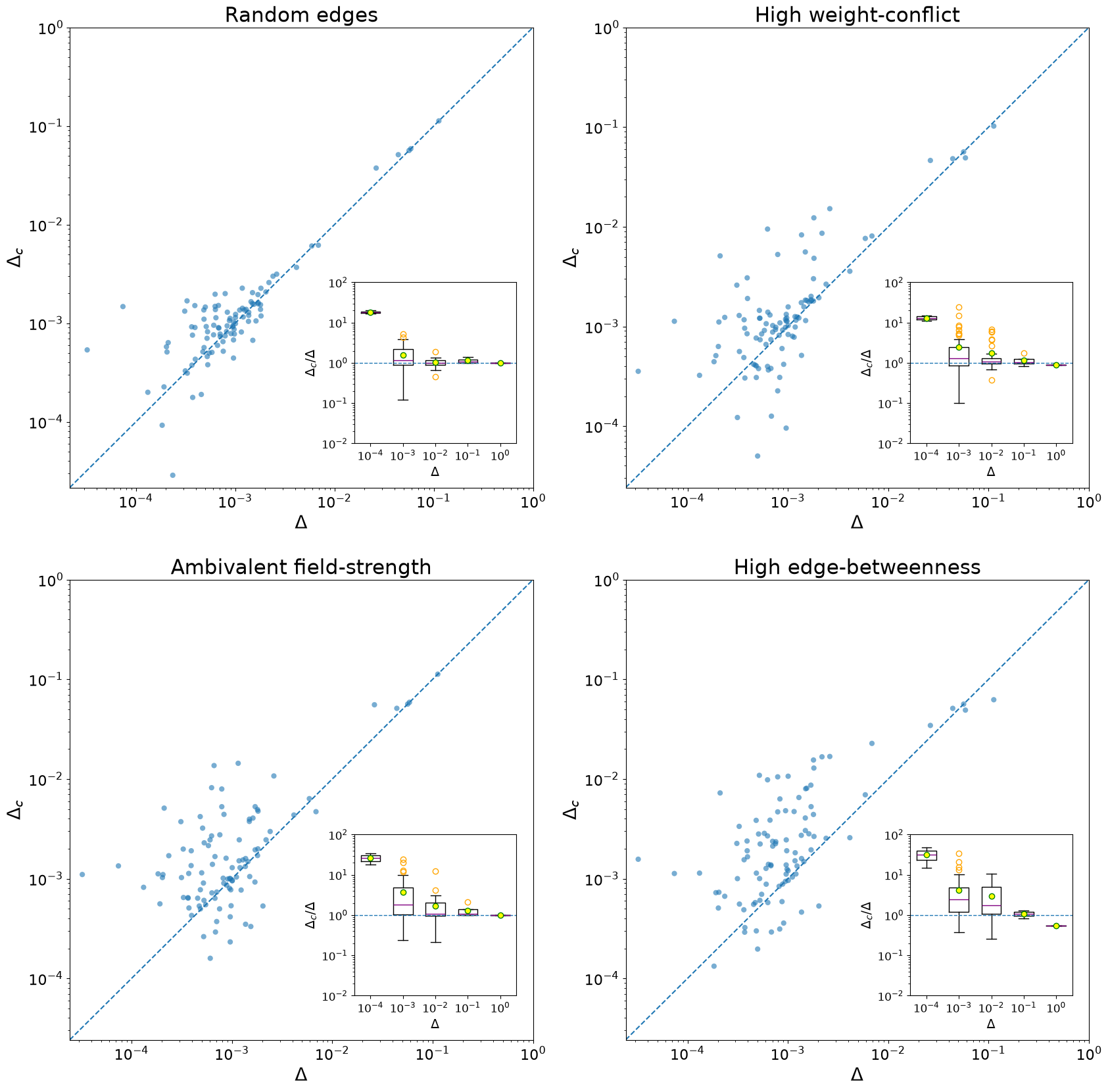}
    \caption{Comparison of the original minimum energy gap $\Delta$ with the minimum gap $\Delta_c$ obtained after introducing three $XX$ terms on $100$ constructed hard MWIS instances, for the four edge-selection strategies: random edges, high weight-conflict, ambivalent field-strength, and high edge-betweenness. Both axes are logarithmic. The dashed line denotes $\Delta_c=\Delta$; points above this line therefore correspond to an increase in the minimum gap from the $XX$ interactions. The insets show the distribution of the improvement ratio $\Delta_c/\Delta$, grouped according to the order of magnitude of the original gap $\Delta$. The purple horizontal line within each box denotes the median and the green circular marker denotes the mean. The lower and upper edges of each box correspond to the first and third quartiles, respectively. Whiskers extend to the most extreme data points lying within $1.5$ times the interquartile range of the corresponding quartile, while points outside this range are shown individually as outliers. In accordance with the notation in~\cite{nutricati2025}, an inset label $10^{-n}$ denotes instances with $10^{-(n+1)} < \Delta \leq 10^{-n}$.}
    \label{fig:planted_hard_MWIS_strategy_comparison}
\end{figure*}

We assume a $XX$ budget of $3$ (i.e., precisely $3$ $XX$ edges), and compare the three topology-constrained strategies in Section~\ref{sec:strategies} with the case without $XX$ interactions. In addition, we also benchmark the results from strategic placement against the random placement of three $XX$ edges for each instance. Fig.~\ref{fig:planted_hard_MWIS_strategy_comparison} summarises the results. Points above the dashed line, defined by $\Delta_c = \Delta$, indicate an enhancement of the minimum gap. Across all strategies, the largest improvements are observed for instances with small original gaps, while the gap values with and without the additional $XX$ terms approach each other as $\Delta$ increases. Random $XX$ edges generally produces more modest enhancements, with most instances remaining comparatively close to the $\Delta_c=\Delta$ line. In contrast, the topology-constrained strategies yield substantially larger improvements in the small-gap regime, with the ambivalent field-strength and high edge-betweenness strategies producing increases of more than an order of magnitude for the smaller-gap instances.

\subsection{Scaling with the $XX$ budget}
\begin{figure*}[t]
    \centering
    \includegraphics[width=0.95\linewidth]{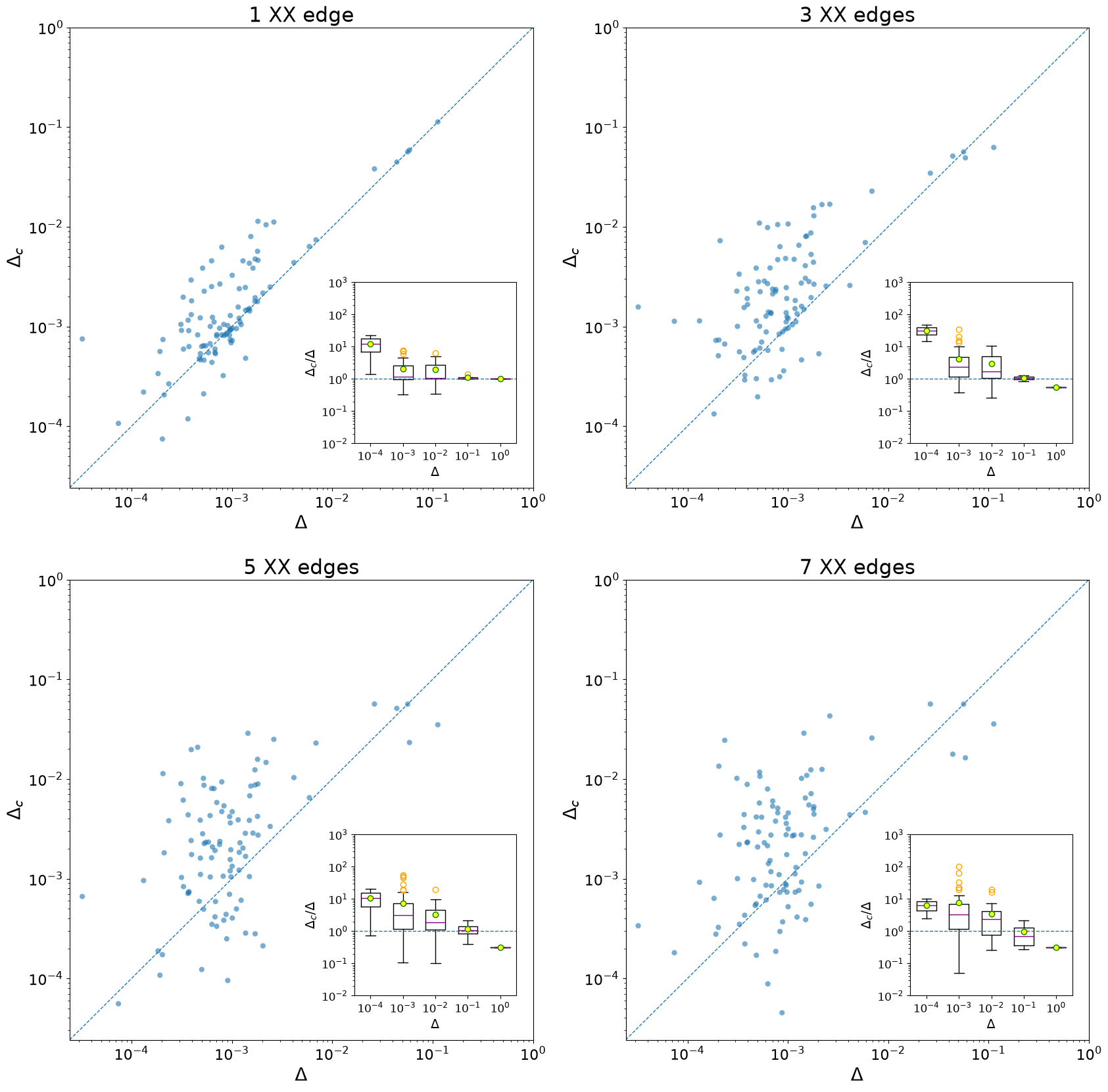}
    \caption{Effect of increasing the $XX$ budget for the high edge-betweenness placement strategy on $100$ constructed hard MWIS instances. The panels show a comparison of the original minimum energy gap $\Delta$ with the minimum gap $\Delta_c$ obtained after introducing $XX$ terms for budgets of $1$, $3$, $5$, and $7$ $XX$ edges. The dashed line denotes $\Delta_c = \Delta$, such that points above the line correspond to an enhancement of the minimum gap. Increasing the $XX$ budget generally improves the minimum gap, particularly for instances with small $\Delta$, although the additional improvement diminishes at larger budgets. The insets show the distribution of the improvement ratio $\Delta_c / \Delta$, grouped according to the order of magnitude of the original gap $\Delta$. The purple line and the green circular marker denote the median and mean, respectively; the box edges indicate the first and third quartiles, the whiskers extend to the most extreme non-outlier values, and orange circles denote outliers more than 1.5 times more than the inter-quartile range. An inset label $10^{-n}$ denotes the interval $10^{-(n+1)} < \Delta \leq 10^{-n}$.}
    \label{fig:planted_hard_MWIS_scaling_with_n_catalysts}
\end{figure*}

To investigate how the performance scales with the $XX$ budget, we fix the placement strategy to high edge-betweenness and increase the number of $XX$ edges from $1$ to $7$. Fig.~\ref{fig:planted_hard_MWIS_scaling_with_n_catalysts} shows that increasing the $XX$ budget generally increases the enhancement of the minimum gap, with the effect again being most pronounced for instances with small original gaps. The improvement is substantial when increasing the budget from a single $XX$  to several $XX$ interactions, as reflected by both the increasing number of points above the $\Delta_c = \Delta$ line and the larger values of $\Delta_c / \Delta$ in the inset distributions. However, the improvement exhibits diminishing returns. Beyond a moderate $XX$ budget, increasing the number of $XX$ edges further produces only marginal additional enhancement. For instances with larger original gaps, $\Delta_c / \Delta$ remains concentrated near unity, indicating that increasing the $XX$ budget provides little additional benefit in this regime.

\subsection{Scaling with the length of connected \texorpdfstring{$XX$}{XX} terms}
\begin{figure*}[t]
    \centering
    \includegraphics[width=0.95\linewidth]{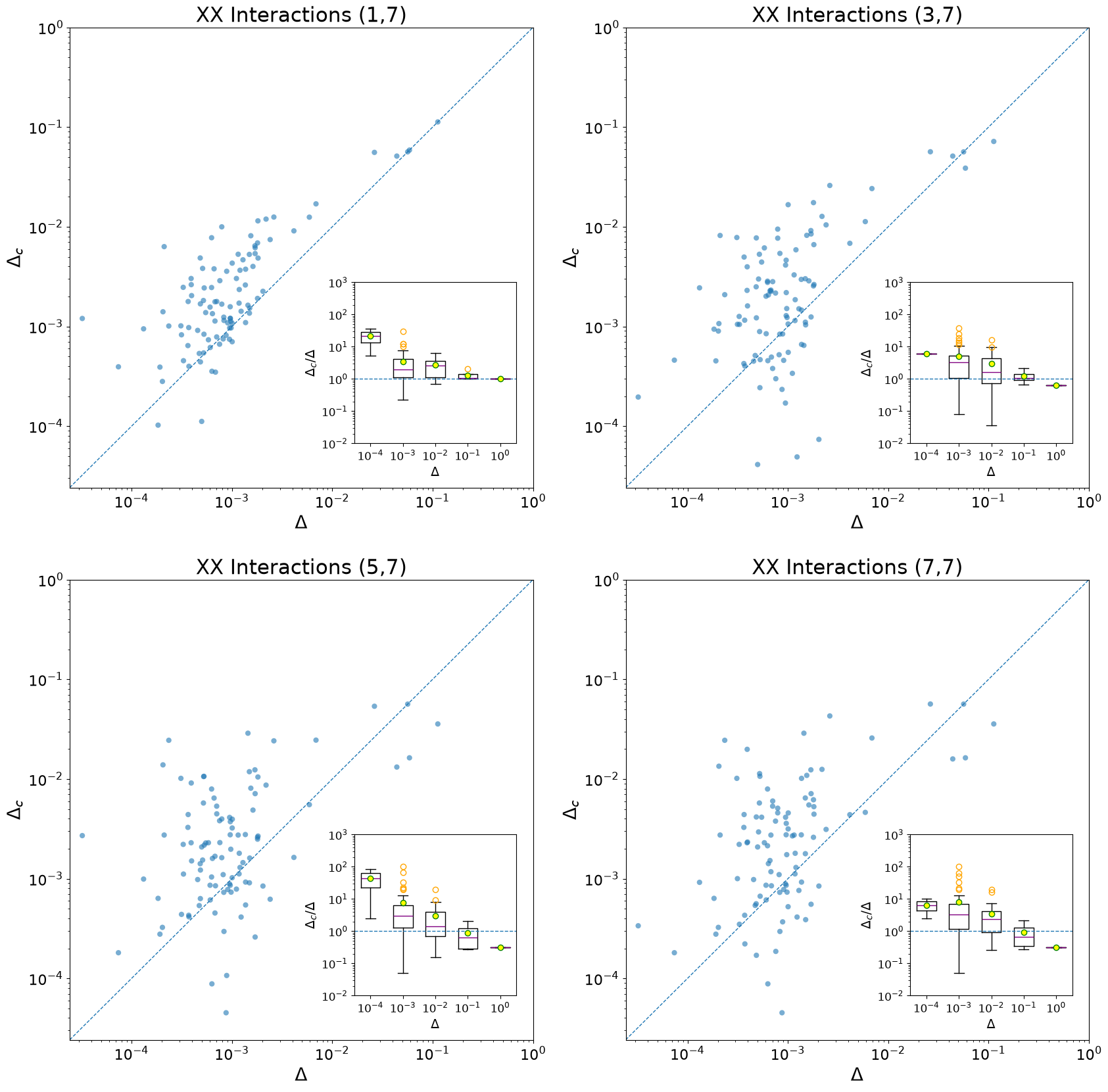}
    \caption{Effect of the maximum connected-component length of the $XX$-interaction graph for the high edge-betweenness placement strategy at fixed $XX$ budget $7$,  on $100$ constructed hard MWIS instances. The four panels correspond to the cases $(1,7)$, $(3,7)$, $(5,7)$, and $(7,7)$, where the first entry denotes the maximum allowed length of a connected $XX$ component and the second denotes the total number of $XX$ edges included. Each panel shows a comparison of the original minimum energy gap $\Delta$ with the minimum gap. The dashed line denotes $\Delta_c = \Delta$, so points above the line correspond to an enhancement of the minimum gap. For the smallest-gap instances, increasing the allowed component length beyond a certain value exhibits diminishing returns and can in some cases reduce the gap enhancement, reflecting the trade-off between increased $XX$ connectivity and the loss of single-$X$ degrees of freedom in the parity-based gadget construction. The insets show the distribution of $\Delta_c / \Delta$, grouped by the order of magnitude of $\Delta$. The purple line and green circular marker denote the median and mean, respectively; the box edges indicate the first and third quartiles, whiskers extend to the most extreme non-outlier values within $1.5$ times the interquartile range, while orange circles denote outliers. An inset label $10^{-n}$ denotes the interval $10^{-(n+1)} < \Delta \leq 10^{-n}$.}
    \label{fig:planted_hard_MWIS_scaling_with_length_of_catalyst_component}
\end{figure*}

To assess the impact of the parity-based gadget construction constraint on the $XX$ graph geometry, we fix the placement strategy to high edge-betweenness and the total $XX$ budget to $7$, and vary the maximum allowed length of a connected $XX$ graph component. As discussed in Section~\ref{sec:ancilla_implementation}, in the parity-based gadget implementation of $XX$ couplings, when several $XX$ edges are connected into a single component, qubits in the interior of that component lose the ability to support independent single-$X$ terms. Fig.~\ref{fig:planted_hard_MWIS_scaling_with_length_of_catalyst_component} therefore probes the trade-off between allowing more extended $XX$ graph structures and preserving local transverse-field degrees of freedom. For the smallest-gap instances, increasing the maximum component length does not lead to a monotonic improvement. While allowing short connected components can retain substantial gap enhancement, the benefit diminishes as longer components are permitted, and in some cases the performance becomes worse. This suggests that the gain obtained from introducing additional connectivity to the $XX$-interaction graph can be offset by the corresponding loss of local transverse-field degrees of freedom.

\section{Numerical analysis on Erd\H{o}s--R\'enyi graph-instances}
\label{sec:ER_instances}
\begin{definition}[Erd\H{o}s--R\'enyi graphs]
    An Erd\H{o}s--R\'enyi graph $G(n,p)$ is a simple undirected graph on the vertex set $V=\{1,\dots,n\}$, in which each unordered pair of distinct vertices $\{i,j\}\subseteq V$ is included as an edge independently with probability $p$.
\end{definition}

Following~\cite{nutricati2025}, we extend our investigation to random Erd\H{o}s--R\'enyi graph instances of the MWIS problem. We consider graphs with $N=10$ vertices and edge probability $p=0.6$. For each graph, the MWIS parameters are sampled independently and uniformly, according to
\begin{equation}
    J_{ij} \sim \mathcal{U}[1,2],
    \qquad
    w_i \sim \mathcal{U}[0,1],
\end{equation}
for every $(i,j)\in E$ and $i\in V$, respectively. The choice $J_{ij}>w_i$ ensures that the energetic penalty associated with selecting adjacent vertices dominates the contribution from any individual vertex weight, so that configurations violating the independence constraint are energetically disfavoured.

\begin{figure}[t]
    \centering
    \includegraphics[width=0.85\linewidth]{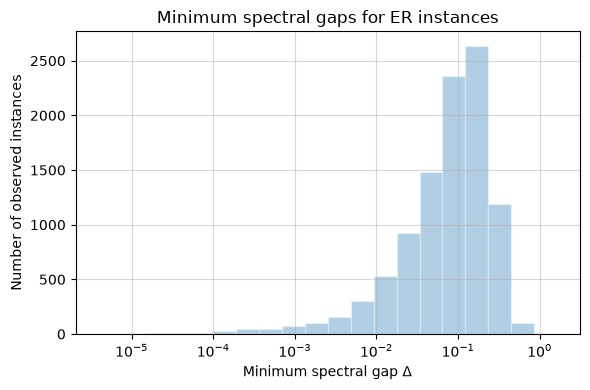}
    \caption{Distribution of minimum spectral gaps $\Delta$ for MWIS instances on random Erd\H{o}s--R\'enyi graphs prior to the introduction of $XX$ interactions. The histogram shows the number of instances within logarithmically spaced bins. The distribution is concentrated between $10^{-2} < \Delta \leq 10^{-0}$, as they do not contain any explicitly engineered hard substructure.}
    \label{fig:distribution_of_ER_delta}
\end{figure}

Fig.~\ref{fig:distribution_of_ER_delta} shows the distribution of minimum spectral gaps $\Delta$ for MWIS instances generated on random Erd\H{o}s--R\'enyi graphs prior to the introduction of $XX$ interactions. Unlike the planted small-gap instances considered in the previous section, the distribution of $\Delta$ is concentrated between $10^{-2} < \Delta \leq 10^{-0}$, as these ER instances do not contain an explicitly engineered hard substructure. They therefore provide a complementary test of whether the $XX$ placement strategies remain effective on generic random graph topologies.

\subsection{Comparison of \texorpdfstring{$XX$}{XX} placement strategies}
\begin{figure*}[t]
    \centering
    \includegraphics[width=0.95\linewidth]{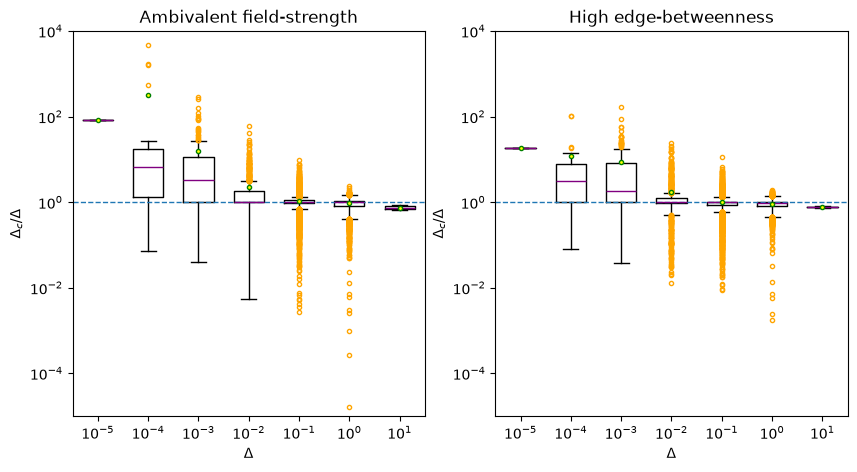}
    \caption{Performance of the ambivalent field-strength and high edge-betweenness placement strategies on $10,000$ random Erd\H{o}s--R\'enyi MWIS instances. For both strategies, the $XX$ budget is fixed to $7$, with connected $XX$ graph components restricted to a maximum length of $3$. The box plots show the distribution of the improvement ratio $\Delta_c/\Delta$, grouped according to the order of magnitude of the original minimum gap $\Delta$. The dashed horizontal line denotes $\Delta_c/\Delta=1$, corresponding to no change in the minimum gap. Both strategies yield little or slightly detrimental effect for large-gap instances, while producing increasingly strong enhancements as $\Delta$ decreases, with the largest improvements occurring for the smallest-gap instances. The purple horizontal line and green circular marker denote the median and mean, respectively; the box edges indicate the first and third quartiles, the whiskers extend to the most extreme non-outlier values, and orange circles denote outliers more than $1.5$ times from the interquartile range. An axis label $10^{-n}$ denotes the interval $10^{-(n+1)}<\Delta\leq10^{-n}$.}
    \label{fig:ER_instance_strategy_comparison}
\end{figure*}

To test whether the $XX$ placement strategies generalise beyond the planted small-gap instances, we generate $10,000$ random Erd\H{o}s--R\'enyi MWIS instances, allowing us to sample a sufficient number of instances with small minimum gaps. We consider the two topology-constrained strategies that performed best on the planted small-gap instances, namely ambivalent field-strength and high edge-betweenness. For both strategies, we fix the $XX$ budget to $7$ and restrict each connected $XX$-interaction graph component to have length at most $3$. Fig.~\ref{fig:ER_instance_strategy_comparison} shows that both strategies retain the same qualitative behaviour observed for the planted small-gap instances. For instances with relatively large original gaps, the addition of $XX$ interactions provides little improvement and can even lead to a modest reduction in the minimum gap, with $\Delta_c / \Delta$ concentrated around or slightly below unity. In contrast, the enhancement becomes increasingly pronounced as the original gap decreases. In the smallest-gap regimes, both strategies produce substantial increases in the minimum gap, often by more than an order of magnitude. This indicates that the topology-constrained placement strategies generalise to random Erd\H{o}s--R\'enyi instances, with their principal benefit concentrated on the hardest, small-gap instances.

\subsection{Comparison with unconstrained \texorpdfstring{$XX$}{XX} placement}
We believe another meaningful question to investigate is whether the gap enhancements reported in~\cite{nutricati2025}, where an $XX$ interaction is introduced for each problem edge, can be reproduced subject to the implementation constraints imposed by the parity-based gadget construction. Rather than selecting the placement of $XX$ interactions according to a scoring criterion, we therefore seek to place $XX$ edges on as many problem edges as possible while respecting the local topology constraints of the implementation.

Since the parity-based gadget construction requires the $XX$-interaction graph to be non-branching, every vertex may have degree at most two. At first glance, the resulting optimisation problem translates nicely to the maximum-cardinality $2$-matching problem.

\begin{definition}[Maximum $2$-matching]
    Let $G=(V,E)$ be a simple undirected graph. A \emph{maximum $2$-matching} is a largest edge set $M\subseteq E$ such that
    $$
        \deg_M(v)\leq 2
        \qquad
        \forall v\in V.
    $$
    where $\deg_M(v)$ denotes the number of edges in $M$ incident on $v$. 
\end{definition}

The maximum $2$-matching problem is a standard special case of $b$-matching and can be reduced in polynomial time to an ordinary matching problem~\cite{schrijver2003}, which may then be solved using Edmonds' blossom algorithm~\cite{edmonds1965}. As this procedure is well established, we do not reproduce it here and instead refer the interested reader to these classical results. For our application, however, the degree constraint alone is insufficient. Under the parity-based gadget construction, vertices lying in the interior of a connected $XX$ graph component lose their independent single-$X$ degree of freedom. We therefore consider a greedy bounded $2$-matching heuristic that iteratively selects candidate problem edges while enforcing both the degree and component-size constraints. Specifically, the candidate edges are first ordered so that edges incident on lower-degree vertices are considered first, with ties broken randomly. The edges are then processed sequentially: an edge is accepted only if neither endpoint already has degree two and if the edge's addition would not cause the resulting connected component to exceed the prescribed maximum number of edges. Connected components and their sizes are tracked dynamically using a union--find data structure, allowing both the merging of distinct path components and the closure of a path into a loop to be checked against the component-size bound. The resulting $XX$ graph is therefore a maximal disjoint union of paths and cycles of bounded size, although the greedy construction is not guaranteed to yield a maximum-cardinality set of feasible edges.

\begin{figure*}[t]
    \centering
    \includegraphics[width=0.95\linewidth]{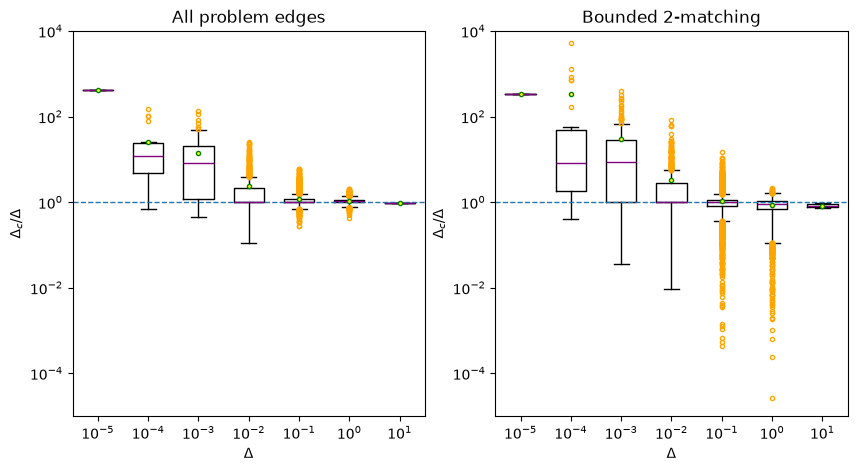}
    \caption{Comparison of the all-problem-edge construction (“All problem edges”) and the topology-constrained bounded $2$-matching construction (“Bounded $2$ matching”), each benchmarked against the case with no $XX$ interactions. In bounded $2$-matching, the lengths of connected $XX$ graph components are additionally constrained to be less than or equal to $3$. The box plots show the distribution of the ratio $\Delta_c/\Delta$, grouped according to the order of magnitude of the original minimum gap $\Delta$. The dashed horizontal line denotes $\Delta_c/\Delta=1$, corresponding to no change in the minimum gap. In both constructions, the largest gap enhancements occur for the smallest-gap instances, while the effect becomes weaker as $\Delta$ increases. The bounded $2$-matching heuristic reproduces the same qualitative improvement on hard instances as the all-problem-edge construction, but on average only placed 7.8 $XX$ edges per instance, compared to 27 per instance on all problem edges. The horizontal line within each box and green circular marker denote the median and mean, respectively; the box edges indicate the first and third quartiles, the whiskers extend to the most extreme non-outlier values, and individual points denote outliers more than $1.5$ times from the interquartile range. An axis label $10^{-n}$ denotes the interval $10^{-(n+1)}<\Delta\leq10^{-n}$.}
    \label{fig:ER_bounded_2_matching_vs_all_problem_edge}
\end{figure*}

To compare the unconstrained and topology-constrained constructions on the same footing, Fig.~\ref{fig:ER_bounded_2_matching_vs_all_problem_edge} benchmarks both the all-problem-edge placement and the bounded $2$-matching heuristic against the no-$XX$ case. In the bounded $2$-matching heuristic, we restrict the connected $XX$ graph components to a maximum length of three. In both panels, the strongest enhancements occur for instances with the smallest original minimum gaps, where $\Delta_c / \Delta$ is often well above unity and can exceed one or two orders of magnitude. This shows that both constructions are able to substantially enlarge the gap precisely in the hard-instance regime. As the original gap increases, however, the improvement decreases and the ratio concentrates around $\Delta_c / \Delta \approx 1$, indicating little effect on easier instances. Comparing the two panels, the bounded $2$-matching heuristic broadly preserves the qualitative advantage of the all-problem-edge construction in the small-gap regime, while exhibiting a wider spread of outcomes and a greater number of instances with $\Delta_c/\Delta<1$ for intermediate and larger gaps. Importantly the bounded $2$-matching heuristic achieves this result with consistently fewer $XX$ edges. On average, the bounded $2$-matching heuristic places 7.8 $XX$ edges per instance, compared to 27 $XX$ edges placed per instance on all problem edges.

Thus, the topology-constrained construction retains the main benefit of introducing $XX$ interactions to hard instances with the need for fewer $XX$ edges, albeit with increased variability relative to the unconstrained placement of $XX$ terms on all problem edges.

\section{Concluding remarks}
\label{sec:conclusion}
In this work we have investigated efficient strategies for placing a limited number of $XX$ edges in quantum annealing for the maximum weight independent set problem, while explicitly taking into account both resource constraints and the topological restrictions arising from an parity-based gadget implementation. We introduced three polynomial-time placement strategies based on high vertex-weight conflict, ambivalent local field-strength, and high edge-betweenness, and evaluated their ability to enhance the minimum spectral gap on both deliberately constructed small-gap instances and random Erd\H{o}s--R\'enyi graphs.

Our numerical results show that carefully selected subsets of $XX$ edges can produce substantial gap enhancements, even when the $XX$ interactions are placed only on a small fraction of the problem edges. The benefit is strongly concentrated in the hardest instances: as the original minimum gap decreases, the relative enhancement obtained from $XX$ placement becomes increasingly pronounced, with improvements of more than an order of magnitude observed for many of the smallest-gap instances. Among the strategies considered, ambivalent field-strength and high edge-betweenness consistently perform particularly well. These qualitative trends persist for random Erd\H{o}s--R\'enyi instances, indicating that the observed improvement is not restricted to the particular structure of the planted small-gap instances.

We further investigated how the performance depends on the available $XX$ budget and on the topology of the $XX$-interaction graph. Increasing the budget initially leads to substantial improvements, but with diminishing returns once a moderate number of $XX$ interactions has been introduced. Likewise, allowing increasingly long connected $XX$ graph components does not necessarily improve the minimum gap. Within the parity-based gadget construction, longer components remove independent single-$X$ degrees of freedom from interior qubits, and the resulting loss of local transverse driving can offset the benefit gained from introducing additional $XX$ connectivity. This demonstrates that the effectiveness of $XX$-term placement depends not only on the number of $XX$ interactions, but also on the geometry of the interaction graph.

Finally, we compared the topology-constrained construction with the unconstrained placement of $XX$ interactions on all problem edges. A greedy bounded $2$-matching construction was able to retain the principal gap enhancement observed for the hardest instances while using substantially fewer $XX$ interactions. This suggests that much of the advantage of dense $XX$ placement can survive under considerably more restrictive and experimentally motivated connectivity constraints.

A key advantage of the standard driver Hamiltonian is the ease with which its ground state can be prepared. Although we have not yet addressed initialisation for the modified driver in Eq.~\eqref{eq:new_driver}, the gadget construction underlying our approach means that the anneal is initialised not in the ground state of the modified driver, but in that of the standard driver in Eq.~\eqref{eq:driver_hamiltonian}, which describes the physical Hamiltonian of the system. Some of the subtleties of what this means for the initial \textit{logical} state are discussed in~\cite{banks2026}. Crucially, since the ground-state of the physical gadget Hamiltonian at the end of the anneal corresponds to the original problem ground-state, adiabatic evolution starting in the ground state of the standard driver provides a route to the optimal solution. 

An important direction for future work is to refine the structure of the parity-based gadget construction itself. In particular, it would be useful to design $XX$-interaction graph geometries that preserve the favourable gap enhancement while minimising the occurrence of higher-order $Z$ interactions and other implementation overheads introduced by the mapping. More generally, jointly optimising the choice of $XX$ edges, connected-component geometry, and ancilla-qubit placements could provide a route towards interaction graph choices that are both spectrally effective and more compatible with realistic quantum-annealing hardware. Additionally, extending these investigations to larger problem sizes and alternative classes of optimisation problems also would further clarify how broadly such constrained $XX$ placement strategies can be applied.

\begin{acknowledgments}
This work is supported by the  QCi3 Hub, funded by EPSRC (grant ref. EP/Z53318X/1).
\end{acknowledgments}

\bibliographystyle{apsrev4-2}
\bibliography{bibliography}

@misc{farhi2000,
    title={Quantum Computation by Adiabatic Evolution}, 
    author={Edward Farhi and Jeffrey Goldstone and Sam Gutmann and Michael Sipser},
    year={2000},
    eprint={quant-ph/0001106},
    archivePrefix={arXiv},
    primaryClass={quant-ph},
}

@article{albash2018,
    title={Adiabatic Quantum Computation},
    author={Albash, Tameem and Lidar, Daniel A.},
    journal={Reviews of Modern Physics},
    volume={90},
    number={1},
    pages={015002},
    year={2018},
    doi={10.1103/RevModPhys.90.015002}
}

@article{nutricati2025,
    title={Enhancing the Energy Gap of Random Graph Problems via {XX}-Catalysts in Quantum Annealing},
    author={Nutricati, Luca A. and Ghosh, Roopayan and Feinstein, Natasha and Bose, Sougato and Warburton, Paul A.},
    journal={Quantum Science and Technology},
    volume={10},
    pages={045010},
    year={2025},
    doi={10.1088/2058-9565/adf2d5}
}

@misc{sales2023,
    title={Adiabatic Quantum Computing for Logistic Transport Optimization}, 
    author={Juan Francisco Ariño Sales and Raúl Andres Palacios Araos},
    year={2023},
    eprint={2301.07691},
    archivePrefix={arXiv},
    primaryClass={quant-ph},
}

@article{gabbassov2022,
    title={Transit facility allocation: Hybrid quantum-classical optimization},
    volume={17},
    DOI={10.1371/journal.pone.0274632},
    number={9},
    journal={PLOS ONE},
    author={Gabbassov, Einar},
    year={2022},
    pages={e0274632}
}

@INPROCEEDINGS{witt2023,
    author={Witt, Arthur and Körber, Christopher and Kirstädter, Andreas and Luu, Thomas},
    booktitle={2023 IEEE International Conference on Quantum Computing and Engineering (QCE)}, 
    title={Tactile Network Resource Allocation Enabled by Quantum Annealing Based on ILP Modeling}, 
    year={2023},
    volume={01},
    pages={670-680},
    doi={10.1109/QCE57702.2023.00082}
}

@ARTICLE{otsuka2022,
    author={Otsuka, Teppei and Li, Aohan and Takesue, Hiroki and Inaba, Kensuke and Aihara, Kazuyuki and Hasegawa, Mikio},
    journal={IEEE Transactions on Vehicular Technology}, 
    title={High-Speed Resource Allocation Algorithm Using a Coherent Ising Machine for NOMA Systems}, 
    year={2024},
    volume={73},
    number={1},
    pages={707-723},
    doi={10.1109/TVT.2023.3300920}
}

@misc{rubiogarcia2022,
    title={Portfolio optimization with discrete simulated annealing}, 
    author={Álvaro Rubio-García and Juan José García-Ripoll and Diego Porras},
    year={2022},
    eprint={2210.00807},
    archivePrefix={arXiv},
    primaryClass={cond-mat.stat-mech},
}

@article{santoro2002,
    author={Santoro, Giuseppe E. and Marto{\v n}{\'a}k, Roman and Tosatti, Erio and Car, Roberto},
    title={Theory of Quantum Annealing of an {Ising} Spin Glass},
    journal={Science},
    volume={295},
    number={5564},
    pages={2427--2430},
    year={2002},
    doi={10.1126/science.1068774},
}

@misc{banks2025,
    title={Gadgets for simulating a non-native $XX$ interaction in quantum annealing}, 
    author={Robert J. Banks and Natasha Feinstein and Roopayan Ghosh and Sougato Bose and P. A. Warburton},
    year={2025},
    url={https://arxiv.org/abs/2503.16663}, 
}

@inproceedings{hagberg2008,
    author={Hagberg, Aric A. and Schult, Daniel A. and Swart, Pieter J.},
    title={Exploring Network Structure, Dynamics, and Function Using NetworkX},
    booktitle={Proceedings of the 7th Python in Science Conference},
    pages={11--15},
    year={2008},
    doi={10.25080/TCWV9851}
}

@manual{networkx_centrality_docs,
    author={{NetworkX Developers}},
    title={Centrality Algorithms --- NetworkX Documentation},
    organization={NetworkX},
    year={2026},
    url={https://networkx.org/documentation/stable/reference/algorithms/centrality.html},
    note={Accessed: 2026-07-24}
}

@article{BRANDES2008136,
    title={On variants of shortest-path betweenness centrality and their generic computation},
    journal={Social Networks},
    volume={30},
    number={2},
    pages={136-145},
    year={2008},
    doi={https://doi.org/10.1016/j.socnet.2007.11.001},
    author={Ulrik Brandes},
}

@article{feinstein2024,
    title={Robustness of diabatic enhancement in quantum annealing},
    doi={10.1088/2058-9565/adab14},
    year={2025},
    volume={10},
    number={2},
    pages={025011},
    author={Feinstein, Natasha and Shalashilin, Ivan and Bose, Sougato and Warburton, P A},
    journal={Quantum Science and Technology},
}

@article{edmonds1965, 
    title={Paths, Trees, and Flowers},
    volume={17},
    DOI={10.4153/CJM-1965-045-4},
    journal={Canadian Journal of Mathematics},
    author={Edmonds, Jack},
    year={1965},
    pages={449–467}
}

@book{schrijver2003,
    title={Combinatorial Optimization: Polyhedra and Efficiency},
    author={Schrijver, Alexander},
    volume={24},
    year={2003},
    publisher={Springer Science \& Business Media}
}

@misc{banks2026,
    title={Enhanced analogue quantum computation with a logically-encoded XX gadget}, 
    author={Robert J. Banks and Natasha Feinstein and Roopayan Ghosh and Sebastian Schulz and Kristel Michielsen and Sougato Bose and P. A. Warburton},
    year={2026},
    eprint={2610.13845},
    archivePrefix={arXiv},
    primaryClass={quant-ph},
}

@article{SekiYuyaandNishimori2012,
    title = {Quantum annealing with antiferromagnetic fluctuations},
    author = {Yuya Seki and Hidetoshi Nishimori},
    year = {2012},
    volume = {85},
    journal = {Physical Review E},
    pages = {051112},
    doi = {10.1103/PhysRevE.85.051112},

}

@article{Hormozi2017,
    title = {Nonstoquastic Hamiltonians and quantum annealing of an Ising spin glass},
    author = {Layla Hormozi and Ethan W. Brown and Giuseppe Carleo and Matthias Troyer},
    year = {2017},
    volume = {95},
    journal = {Physical Review B},
    pages = {184416},
    doi = {10.1103/PhysRevB.95.184416},
}

@article{Albash2019,
    title = {Role of nonstoquastic catalysts in quantum adiabatic optimization},
    author = {Tameem Albash},
    year = {2019},
    volume = {99},
    journal = {Physical Review A},
    pages = {042334},
    doi = {10.1103/PhysRevA.99.042334},
}

@article{Mehta2021,
    title={Quantum annealing with trigger Hamiltonians: Application to 2-satisfiability and nonstoquastic problems},
    author={Vrinda Mehta and Fengping Jin and Hans De Raedt and Kristel Michielsen},
    year={2021},
    volume={104},
    journal={Physical Review A},
    publisher={American Physical Society},
    pages = {032421},
    doi = {10.1103/PhysRevA.104.032421},
}

@article{feinstein2024-xx-catalyst,
    title={Effects of $XX$ catalysts on quantum annealing spectra with perturbative crossings},
    author={Natasha Feinstein and Louis Fry-Bouriaux and Sougato Bose and P. A. Warburton},
    journal={Physical Review A},
    volume={110},
    pages={042609},
    year={2024},
    publisher={American Physical Society},
    doi = {10.1103/PhysRevA.110.042609},
}

@article{ghosh2024,
    title={Enhancement of quantum annealing via n-local catalysts}, 
    author={Roopayan Ghosh and Luca A. Nutricati and Natasha Feinstein and P. A. Warburton and Sougato Bose},
    year={2026},
    journal={Physical Review Research},
    volume={8},
    pages={013301},
    doi = {10.1103/hnzj-j7kt},
}

@article{Headley_2026,
    title={Non-perturbative topological gadgets for many-body coupling},
    volume={11},
    DOI={10.1088/2058-9565/ae6fe2},
    journal={Quantum Science and Technology},
    author={Headley, David and Chancellor, Nicholas},
    year={2026},
    month={June}, 
    pages={035010} }

@misc{zhang2023tunableinductivecouplerhigh,
    title={Tunable inductive coupler for high fidelity gates between fluxonium qubits}, 
    author={Helin Zhang and Chunyang Ding and D. K. Weiss and others},
    year={2023},
    eprint={2309.05720},
    archivePrefix={arXiv},
    primaryClass={quant-ph},
}

@article{PhysRevApplied.13.034037,
    title={Demonstration of a Nonstoquastic Hamiltonian in Coupled Superconducting Flux Qubits},
    author={Ozfidan, I. and Deng, C. and Smirnov, A.Y. and others},
    journal={Phys. Rev. Appl.},
    volume={13},
    issue={3},
    pages={034037},
    numpages={15},
    year={2020},
    doi={10.1103/PhysRevApplied.13.034037},
}

@article{Kerman_2019,
    doi={10.1088/1367-2630/ab2ee7},
    year={2019},
    volume={21},
    number={7},
    pages={073030},
    author={Kerman, Andrew J},
    title={Superconducting qubit circuit emulation of a vector spin-1/2},
    journal={New Journal of Physics},
}

@misc{dattani2019quadratizationdiscreteoptimizationquantum,
      title={Quadratization in discrete optimization and quantum mechanics}, 
      author={Nike Dattani},
      year={2019},
      eprint={1901.04405},
      archivePrefix={arXiv},
      primaryClass={quant-ph},
}

\end{document}